\documentclass[twocolumn,prb]{revtex4}
\usepackage{epsfig}
\usepackage{graphicx}
\usepackage{amsmath}
\usepackage{amssymb}

\begin{document}


\title{A Two-Region Diffusion Model for Current-Induced Instabilities of Step
Patterns on Vicinal Si(111) Surfaces}
\author{Tong Zhao$^1$ and John D. Weeks$^{1,2}$}
\affiliation{$^1$Institute for Physical Science and Technology, University
of Maryland, College Park, Maryland 20742}
\affiliation{$^2$Department of Chemistry and Biochemistry,
University of Maryland, College Park, Maryland 20742}

\date{\today }

\begin{abstract}
We study current-induced step bunching and wandering instabilities with
subsequent pattern formations on vicinal surfaces. A novel two-region
diffusion model is developed, where we assume that there are different
diffusion rates on terraces and in a small region around a step, generally arising
from local differences in surface reconstruction. We determine the steady
state solutions for a uniform train of straight steps, from which step
bunching and in-phase wandering instabilities are deduced. The
physically suggestive parameters of the two-region model are then mapped
to the effective parameters in the usual sharp step models.
Interestingly, a negative kinetic coefficient results when the diffusion
in the step region is faster than on terraces. A consistent physical
picture of current-induced instabilities on Si(111) is suggested based
on the results of linear stability analysis. In this picture the step
wandering instability is driven by step edge diffusion and is not of the
Mullins-Sekerka type. Step bunching and wandering patterns at longer
times are determined numerically by solving a set of coupled equations
relating the velocity of a step to local properties of the step and its
neighbors. We use a geometric representation of the step to derive a
nonlinear evolution equation describing step wandering, which can
explain experimental results where the peaks of the wandering steps
align with the direction of the driving field.  
\end{abstract}

\maketitle 

\section{Introduction}

Steps on vicinal surfaces exhibit many different instabilities in the
presence of non equilibrium driving forces. Of particular interest to us
here are the current-induced instabilities on Si surfaces that were first
discovered by Latyshev {\it et al.} \cite{Latyshev89} in 1989. After
resistive heating of a vicinal Si(111) surface with a step-down direct
current at temperature around $900^{\circ }C$, they observed the formation
of closely packed step bunches separated by wide step-free terraces. The
the uniform step train remained stable on heating with a step-up current. This
instability has a mysterious temperature dependence, \cite
{Si111bchexp_Homma1,Si111bchexp_Homma2,Si111bchexp_Williams,Si111bchexp_Yamag,Si111bchexp_Yang}
with three temperature ranges between $830^{\circ }C$ and $1300^{\circ }C$
where the unstable current direction reverses.

Furthermore, recent experiments \cite
{Si111wanexp_Degawa1,Si111wanexp_Degawa2,Si111wanexp_Degawa3} in temperature
range II (about $1050^{\circ }C$ to $1150^{\circ }C$) have shown that after
heating for several hours with a step-down current, the initially uniform
steps exhibit a novel {\em wandering instability} with finite wavelength
in-phase sinusoidal undulations in their positions. When the current is
directed at an angle to the average step direction, the undulations are
continuously distorted by the field until finally all the peaks point in the
direction of the field. \cite{Si111wanexp_PKTN}

These instabilities likely arise from a complex interplay between the driven
diffusion of adatoms induced by the electric field ${\bf E}$ and their
attachment/detachment kinetics at steps, which serve as sources and sinks of
adatoms. (Island formation is not important in the temperature regimes we
consider.) Adatoms are believed to acquire a small effective charge $z^{*}e,$
which includes both electrostatic and ``wind-force'' contributions arising
from scattering of charge carriers, and thus experience a force ${\bf F}
=z^{*}e{\bf E}$ that biases their diffusive motion. \cite{GBCF_Stoyanov}
Typically ${\bf E}$ has a magnitude of about $5\sim 10{\rm V/cm}$ and $z^{*}$
is of the order of $10^{-3}-10^{-1}$ for Si. \cite
{Si111bchexp_Williams,zstarmag_Fu}

Most theoretical methods are based on a generalization
of the approach taken by Burton, Cabrera and Frank (BCF), \cite{BCF_classic}
where one considers field driven diffusion of adatoms on terraces, with boundary
conditions at the steps, viewed as line sources and sinks. We will generally
refer to these generalized BCF models as {\em sharp step models}. Surface
reconstruction typically seen on semiconductor surfaces clearly has
important effects on the movement of adatoms on terraces and may well affect
the attachment kinetics at steps. While it is relatively simple to take
account of reconstruction on terrace diffusion by changing the diffusion
constant, it is much less clear how it should be incorporated into the
boundary conditions at the step edges.

Many different boundary conditions have been proposed, incorporating, e.g.,
asymmetric attachment-detachment barriers,\cite{ESclassic_1,ESclassic_2}
periphery diffusion along a step,\cite{KESE_PierreL} permeable steps,\cite
{Si111bchthy_Stoyanov,Si111bchthy_Sato} and field-dependent kinetic
coefficient,\cite{Si111bchthy_Suga} and researchers have shown that
different combinations can give results that can agree with some experiments
on current-induced step bunching. However, a general understanding of the
physics leading to the sharp step boundary conditions and how they are
affected by reconstruction and the external field is far from clear.

We discuss here a simple model incorporating the key physical features of
driven diffusion and surface reconstruction. It can provide a consistent
explanation of many experimental results on both Si(111) and Si(001)
surfaces in terms of a few effective parameters. The model also provides a
physically suggestive way of interpreting sharp step boundary conditions,
showing how the effective parameters in continuum models can be related to
kinetic processes on vicinal surfaces. \cite{short2region}

In the following, we will set up the basic two-region diffusion model and
then examine the non-equilibrium steady state (NESS) solutions. Step
bunching and wandering regimes are discussed, and combined to provide a
coherent scenario for the complicated Si(111) electromigration experiments.
A mapping to the generalized BCF model is presented next that sheds light on
the sharp step boundary conditions. Finally, we study the long time and
nonlinear behavior of these instabilities and find some intriguing patterns
that resemble many features seen in real experiments on Si(111) surfaces. An
alternate derivation of the basic equations and applications to the
different instabilities seen in Si(001) surfaces is given elsewhere. \cite
{hopping_Zhao}

\begin{figure}[tbp]
\includegraphics[width=76mm,height=57mm]{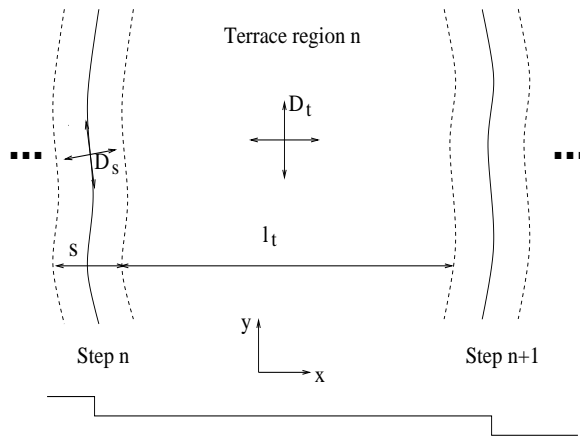}
\caption{The upper part of the figure shows a 2D schematic view of the
vicinal surface composed of different reconstruction regions on terraces and
near steps, separated by dashed lines. In this paper, we assume
that the step reconstruction with a fixed width $s$ always follows the
motion of the step (solid line). The lower part of the figure shows a
corresponding 1D side view that illustrates our coordinate system.}
\label{FIG:tworegion_schm}
\end{figure}

\section{Two-Region Diffusion Model and Steady State Solutions}

It is well known that the dangling bonds at semiconductor surfaces quite
generally rearrange to form characteristic surface reconstructions. We
expect a different local rearrangement of bonds in the vicinity of a step,
which itself represents an additional source of dangling bonds. Clearly this
reconstruction can directly influence surface mass transport and hence
possible instabilities. Standard boundary conditions in the continuum sharp
step model may include some effects of surface reconstruction in special
cases. For example, Liu and Weeks \cite{Si111bchthy_Liu} interpreted
electromigration experiments in the lowest temperature regime of Si(111)
using attachment/detachment limited kinetics, and argued that the attachment
barriers could arise from a local reconstruction of the dangling bonds at a
step edge. However, it is not clear how this picture should be modified at
higher temperatures.

Steps differ fundamentally from terraces by serving as sources and sinks for
adatoms. In the classical BCF picture it was assumed that the local
equilibrium concentration of adatoms at a step is maintained even in the
presence of nonequilibrium driving forces. In addition the rates of various
mass transport processes near steps can differ from kinetic processes on
terraces, e.g., because of differences in local surface reconstructions. The
kinetic coefficients in generalized BCF models try to take both features of
steps into account in an effective way.

Our approach here is to consider a more detailed description where both
features are treated separately in the simplest possible way. We then
obtain the relevant sharp step boundary condition by an appropriate
coarse-graining. To that end, we assume that an atomic step has sufficient
kink sites to maintain a local equilibrium concentration of adatoms as in
the classical BCF picture. Reconstruction is taken into account by assuming
that the atomic step is surrounded by a {\em step region} where adatoms
undergo effective diffusion with a diffusion constant $D_{s}$ that can
differ from $D_{t}$, the value found on terraces.

Here we use the simplest realization of this idea, where the reconstruction
is assumed to occur fast relative to step motion, so that the step region
moves with the atomic step and has a fixed width $s$ of a few lattice
spacings at a given temperature. Thus a uniform vicinal surface can be
viewed as an array of repetitive two-region units, made up of the $n$th step
region and its neighboring lower terrace region. We assume that straight
steps extend along the $y$ direction and that the step index increases in
the step-down direction, defined as the positive $x$ direction, asde
schematically shown in Fig.\ \ref{FIG:tworegion_schm}.

The adatoms undergo driven diffusion from the electric field. The biased
diffusion flux of adatoms with density $c$ takes the form: 
\begin{equation}
{\bf J}_{\alpha }=-D_{\alpha }\nabla c_{\alpha }+D_{\alpha }\frac{{\bf F}}{
k_{B}T}c_{\alpha },  \label{eq:genflux}
\end{equation}
where $\alpha =\left( t,s\right) $ indicates the terrace or step region and $%
D_{\alpha }$ is the diffusion constant in the corresponding region, which
here is taken to be isotropic for simplicity. We also assume that the
effective charge is the same in both regions and ignore the small effects of
step motion on the steady state adatom density field, since the direct
field-induced adatom drift velocity is generally very much larger than the
net velocity of the steps (driven by free sublimation in real experiments)
even at high temperatures. \cite{note_vcheck}

In many studies of step dynamics, because the separation of their respective
time scales, it suffices to solve the diffusion problem with fixed step
positions and then balance the fluxes locally at a step to determine its
motion. This is often called the {\em quasi-stationary} approximation, and
it will be adopted throughout this paper. Thus the static diffusion problem
is simply 
\begin{equation}
\nabla \cdot {\bf J_{\alpha }}=0  \label{eq:gendiff}
\end{equation}
in each region, along with continuity of $c$ and ${\bf J}$ at fixed
boundaries between terrace and step regions. The normal velocity of the step
region is given by mass conservation locally at an infinitesimal portion of
the step region 
\begin{equation}
v_{n}\Delta c=\left[ {\bf J}_{t}^{-}-{\bf J}_{t}^{+}\right] \cdot \hat{n}
-\int_{s}\partial _{\tau }\left[ {\bf J}_{s}\cdot \hat{\tau}\right] .
\label{eq:genvn}
\end{equation}
Here ${\bf J}_{t}^{\pm }$ denote diffusion fluxes in the front and back
terraces respectively and $\Delta c$ is the difference of the areal density
of the two phases --- the solid phase and the 2D adatom gas phase. For
simplicity, we take a simple cubic lattice, so that $\Delta c\approx
1/\Omega =a^{-2}$, where $a$ is the lattice parameter. The last term in
Eq.\ (\ref{eq:genvn}) represents the contribution from diffusion flux in the
step region parallel to the step, where $\tau $ denotes the arc length.

\begin{figure}[tbp]
\includegraphics[width=76mm,height=57mm]{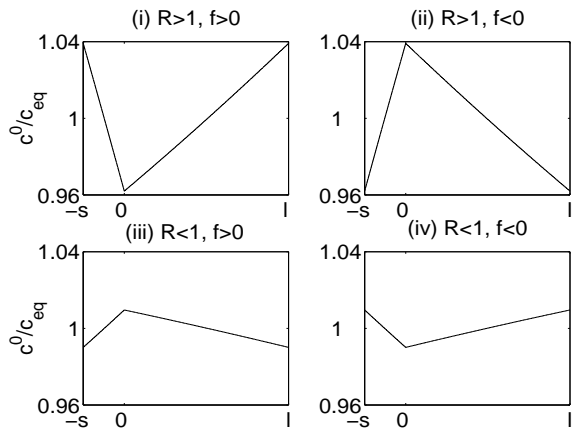}
\caption{Plot of concentration profiles according to Eq.(\ref
{eq:expprofile}) with model parameters. $R=10$ for (i) and (ii), $R=0.1$ for
(iii) and (iv); $\left|fa\right|=0.01$ in all cases.}
\label{FIG:chemgrad_exp}
\end{figure}

Eqs. (\ref{eq:genflux}-\ref{eq:genvn}) define the two-region diffusion
model. We first consider the NESS solution corresponding to a 1D uniform
step train. In this case, the step normal direction coincides with the $x$
direction on terraces, and thus parallel or tangential diffusion in the step
region plays no role here. The NESS concentration profile (denoted by a
superscript '$0$') in a two-region unit is easily obtained by solving Eq.\ (%
\ref{eq:gendiff}) in both regions subject to continuity of concentration and
fluxes at the boundaries and is given by 
\begin{equation}
\begin{array}{cl}
c_{s}^{0} & =C\left[ R+ 
{\displaystyle {\left( 1-R\right) \left( e^{fl_{t}}-1\right)  \over e^{fl_{t}}-e^{-fs}}}%
e^{fx}\right] \\\
c_{t}^{0} & =C\left[ 1- 
{\displaystyle {\left( 1-R\right) \left( 1-e^{-fs}\right)  \over e^{fl_{t}}-e^{-fs}}}%
e^{fx}\right] ,
\end{array}
\label{eq:expprofile}
\end{equation}
Here 
\begin{equation}
R\equiv D_{t}/D_{s}  \label{eq:Rdef}
\end{equation}
is one of the key dimensionless parameters that describes the relative
diffusion rates of adatoms on terraces and in the normal direction of step
regions, $f\equiv {\bf F}\cdot \hat{x}/k_{B}T$ has a dimension of inverse
length and characterizes the strength of the external field, and $l_{t}$ is
average terrace width in the steady state. $C$ is a constant to be
determined shortly. 

Evidently, it is the interplay between the external electric field and
changes in the local diffusion rates, characterized by various combinations
of the two parameters $f$ and $R$, that causes the intriguing instabilities.
With the electric field perpendicular to the step region, altogether there
are four types of steady state adatom concentration profiles with different
combinations of parameters $f$ and $R$, as shown in Fig.\ \ref
{FIG:chemgrad_exp}. In the absence of sublimation, the concentration
profiles we obtain here are completely driven by the external field. By
taking the limit $f\rightarrow 0$ in Eq.\ (\ref{eq:expprofile}), one should
recover the equilibrium concentration (denoted as $c_{eq}$) on the entire
surface. This fixes the constant in Eq.\ (\ref{eq:expprofile}) as 
\begin{equation}
C=c_{eq}\left( l_{t}+s\right) /\left( l_{t}+Rs\right) .  \label{eq:Cvalue}
\end{equation}

Moreover, the constant flux at NESS can be written as 
\begin{equation}
J_{0}(l)=D_{t}c_{eq}f\frac{l}{l+(R-1)s},  \label{eq:JNESS}
\end{equation}
where 
\begin{equation}
l\equiv l_{t}+s  \label{eq:ldef}
\end{equation}
is the distance between the centers of two adjacent step regions in a
uniform step train. 
Note that the NESS concentration profile of adatoms given by Eq.\ (\ref
{eq:expprofile}) reduces to a constant on the entire surface in presence of
the field if the diffusion in the normal step direction is the same as
terrace diffusion, i.e., when $R=1$.

\section{Step Bunching and Wandering Instabilities}

In this section, we study the stability of the NESS solutions. In
particular, the physical origins of both step bunching and wandering
instabilities are qualitatively discussed.

\subsection{Step Bunching Instability}

A common feature of all NESS profiles shown in Fig.\ \ref{FIG:chemgrad_exp}
is that adatom concentration gradients build up in both terrace and step
regions. Under experimentally relevant conditions the field is sufficiently
weak that $fs<fl_{t}\ll 1$ and {\em linear} concentration (or chemical
potential) gradients form. It is then easy to see that the local equilibrium
boundary condition $c=c_{eq}$ in the center of the step region holds automatically
by symmetry. In the qualitative picture of step bunching
discussed by Liu and Weeks, \cite{Si111bchthy_Liu} a positive terrace
concentration gradient (induced in their model by a step-down current with
an attachment barrier at a sharp step edge) leads to step bunching. The
steady state profile they analyzed leading to step bunching in temperature
regime I is very similar to case (i) in Fig. \ref{FIG:chemgrad_exp}. This
corresponds in the two-region model to a step-down field with slower
diffusion in the step region, in agreement with an intuitive picture of a
step barrier.

Moreover, it is clear that profile (iv) is qualitatively the same as (i).
Hence we expect that the steady state (iv), corresponding to {\em faster}
diffusion in the step region with a {\em step-up} field, also undergoes a
bunching instability. A hopping model with these features was studied by
Suga {\it et al.}\cite{Si111bchthy_Suga} by computer simulations, and indeed
they observed a bunching instability.

To understand the bunching of straight steps it is useful to consider a 1D
version of Eq.\ (\ref{eq:genvn}): 
\begin{equation}
v_{n}=\Omega \left[ J_{0}\left( l_{n-1}\right) -J_{0}\left( l_{n}\right)
\right] ,  \label{eq:vnbch}
\end{equation}
where the 1D flux $J_{0}$ as given by Eq.\ (\ref{eq:JNESS}) now depends on
the local terrace widths. Consider a small deviation $\delta
x_{n}=\varepsilon _{n}e^{\omega _{1}t}$ for $n$th step from the NESS, where $%
\varepsilon _{n}\equiv \varepsilon e^{in\phi }$ and $\phi $ is the phase
between neighboring steps. Then the step will move as a result of the
unbalanced fluxes induced by changing width of the terrace in front $%
l_{n}=l+\varepsilon _{n}\left( e^{i\phi }-1\right) $ and back $
l_{n-1}=l+\varepsilon _{n}\left( 1-e^{-i\phi }\right) $. The amplification
rate $\omega _{1}$ is given by $\omega _{1}=v_{n}/\varepsilon _{n}$, and
substituting into Eq.\ (\ref{eq:vnbch}) gives 
\begin{equation}
\begin{array}{cl}
\omega _{1} & =-2\Omega D_{s}%
{\displaystyle {dJ_{0}\left( l\right)  \over dl}}%
(1-\cos \phi ) \\ 
& =2\Omega D_{t}c_{eq}^{0}%
{\displaystyle {f(R-1)s \over [l+(R-1)s]^{2}}}%
\left( 1-\cos \phi \right) .
\end{array}
\label{eq:bnch2r}
\end{equation}
Clearly, step bunching occurs when $f\left( R-1\right) >0$, corresponding to
two different regimes discussed above, and in both cases the most unstable
mode is a step pairing instability with $\phi =\pi $.

\subsection{Step Wandering Instability}

\label{sec:stepwandering}

The 1D NESS concentration profiles also provide important insights into step
wandering, which is essentially a 2D phenomenon. It is clear that the
concentration gradient on the terraces in cases (i) and (iv) can drive a
step wandering instability. The monotonically increasing terrace chemical
potential tends to make a forward bulging part of a step move even faster,
as was first demonstrated for vicinal surfaces by Bales and Zangwill. \cite
{MBEthy_Bales} This is the essence of the classic Mullins-Sekerka
instability. \cite{MS_classic1,MS_classic2} However, as shown above, these
same profiles lead to 1D step bunching, which tends to suppress the
wandering instability. Moreover, this mechanism cannot explain the behavior
in regime II of Si(111) where wandering and bunching occur for {\em different%
} current directions.

\begin{figure}[tbp]
\includegraphics[width=76mm,height=57mm]{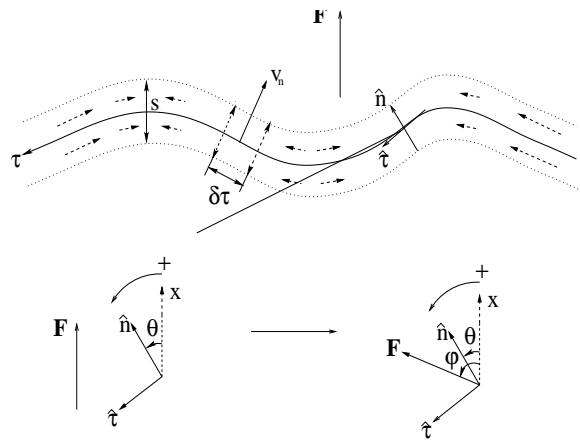}
\caption{A geometrical view of a single wandering step region. The dashed
arrows inside the step region schematically shows the driven flux that is
parallel to the step for a step-down ($x$ direction) field. The lower right
corner shows the case when the field is at an angle $\varphi$ off the $x$-axis.}
\label{FIG:wanderschm}
\end{figure}

The fact that this step wandering cannot be of the Mullins-Sekerka type
driven by terrace gradients suggests that it may originate from mass
transport in the {\em step} region. Let us focus on a single 2D step region,
as in Fig.\ \ref{FIG:wanderschm}. In this case, it is convenient to describe
the step region using curvilinear coordinates set up by the local normal and
tangential directions of the step. For a long wavelength step fluctuation
with wavenumber $q$ there exists a nonzero component of the field in the
tangential direction, which induces a driven flux along the step
proportional to $fq^{2}$. For a step-down field $\left( f>0\right) $, this
driven flux is destabilizing since it tends to transport mass from
``valleys'' to forward-bulging ``hills''. On the other hand, the stabilizing
flux due to the curvature relaxation is proportional to $\Gamma q^{4}$,
where $\Gamma $ is an effective capillary length in the step region. The
competition between these two terms results in a finite wavelength linear
instability, occurring on a length scale of order $\xi $, where 
\begin{equation}
\xi \equiv \sqrt{\Gamma /\left| f\right| }.  \label{def:xi}
\end{equation}

In principle this new wandering instability could arise in cases (i) and
(iii) of Fig.\ \ref{FIG:chemgrad_exp} where there is a step-down field.
However step bunching also occurs for case (i). Only case (iii) with $f>0$
and faster diffusion in the step region ($R<1$) is free of step bunching,
and thus capable of explaining experiments in Regime II of Si(111). In the
next section we show that these qualitative conclusions are in agreement
with a more detailed analysis based on a mapping of the two-region model to
an equivalent sharp step model.

\section{Mapping to a Generalized BCF Model}

In this section we show how the two-region model can be used to generate the
appropriate sharp step boundary conditions by a mapping to a generalized BCF
model.

\begin{figure}[tbp]
\includegraphics[width=76mm,height=57mm]{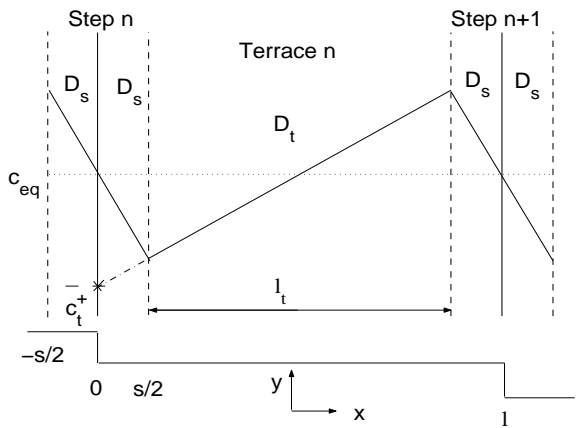}
\caption{Shown is a highly exaggerated profile for a downhill force and
slower diffusion in the step region. Also illustrated with the dashed-dot
line is the extrapolation of the terrace profile to the center of the step
region, thus determining the parameter $\bar{c}_{t}^{+}$ in Eq.\ (\ref
{eq:BCFbc}). The lower part of the figure gives a side view of sharp
equilibrium steps and their associated step regions.}
\label{FIG:extrapolation}
\end{figure}

The general continuum boundary condition in the sharp-step model assumes
small deviations from local equilibrium and introduces linear {\em kinetic
coefficients} $k_{\pm }$ to relate $\bar{c}_{t}^{+}$ (or $\bar{c}_{t}^{-}$),
the limiting lower (or upper) terrace adatom density at the step edge, to
the associated terrace adatom flux into the step. To
linear order in the field this gives rise to the
standard sharp step boundary condition: 
\begin{equation}
\pm D_{t}\left[ \nabla c_{t}-fc_{t}\right] _{\pm }=k\left(
c_{t}-c_{eq}\right) _{\pm }.  \label{eq:BCFbc}
\end{equation}
Here $k$ is the corresponding sharp step kinetic coefficient, which is
symmetric in this case.

A natural way of relating the NESS solutions of the two-region model to those
of sharp step model is to {\em extrapolate} the terrace concentration profile
to the center of the step region. This corresponds to a physical
coarse-graining where the step region has negligible width when compared
to the terrace widths.
The use of extrapolation to relate the parameters in discrete
and continuum models is well known in other interface applications. \cite
{extrapolation} We use Eq.\ (\ref{eq:expprofile}) to evaluate the
gradient, and identify $c_{t}^{\pm }$ as the extrapolated value of
terrace concentration at the atomic step, as illustrated in
Fig.\ \ref{FIG:extrapolation}. Substituting into Eq.\ (\ref{eq:BCFbc}),
to lowest order in the field we find that 
\begin{equation}
d\equiv \frac{D_{t}}{k}=\frac{1}{2}(R-1)s.  \label{eq:mapk}
\end{equation}
Note that the terrace width $l$ in the sharp step model is naturally related
to the two-region width $l_{t}$ by $l=l_{t}+s$, as in Eq.\ (\ref{eq:ldef}).
Here $d$ is often referred to as the attachment-detachment length.

Equation (\ref{eq:mapk}) gives a mapping of the parameters in
the simplest two-region model to those of a generalized BCF model.
When $R>1$ (faster diffusion in the
terrace region), $k$ is positive, which leads to a
bunching instability for a step-down current.
When $R=1$ (the diffusion rate is the same in both
regions), $k$ goes to infinity, which forces $c_{t}^{\pm }$ in Eq.\ (\ref
{eq:BCFbc}) to equal $c_{eq},$ corresponding to local equilibrium with no
instability. When $R<1$ (diffusion is faster in step regions than in terrace
regions), $k$ becomes {\em negative}, which leads to step bunching by a
step-up current together with step wandering by a step-down current.

The possibility of a negative kinetic coefficient, or equivalently a
negative $d$, was first suggested in the work of Politi and Villain, \cite
{ESMBE_Politi} though with no derivation or discussion of any physical
consequences. Note that even though the derivation given here considers
a terrace concentration profile obtained by electromigration,
Eq.\ (\ref{eq:mapk}) is a general result that is independent of the field. In
a related work,\cite{hopping_Zhao} we derive sharp step boundary conditions
by considering a discrete hopping model with different hopping rates in two
regions but without the field, and again obtain Eq.\ (\ref{eq:mapk}).

\begin{table*}

\caption{Linear Stability Results}

\begin{tabular}{|c|c|c|}
\hline 
&
$d>0$ $(R>1)$&
$d<0$ $(R<1)$ \tabularnewline
\hline 
$f>0$&
$\begin{array}{c}
\textrm{Bunching with maximum mode $\phi=\pi$}\\
\textrm{Wandering with maximum mode $\phi=0$}\end{array}$&
 Wandering with maximum mode $\phi=0$  \tabularnewline
\hline 
$f<0$&
 Linearly stable&
$\textrm{Bunching with maximum mode $\phi=\pi$}$ \tabularnewline
\hline
\end{tabular}\label{Table:LSR}
\end{table*}

\section{Linear Stability Analysis}

With the mapping defined by Eq.\ (\ref{eq:mapk}), the linear stability
analysis can be performed using a standard sharp step model, with parameters
obtained from the physically suggestive two-region model. The general result
is presented in the appendix, and here we concentrate on the resulting
stability in the weak field ($fl\ll 1$) and long wavelength ($ql\ll 1$)
limit. The real part of the stability function can be written as 
\begin{equation}
\omega _{r}=\omega _{1}\left( f,\phi \right) +\omega _{2}\left( q,f,\phi
\right) ,  \label{eq:stblygen_bcf_sum}
\end{equation}
where 
\begin{equation}
\omega _{1}=\Omega D_{t}c_{eq}^{0}\frac{4df}{\left( l+2d\right) ^{2}}\left(
1-\cos \phi \right) ,  \label{eq:bch_bcf_sum}
\end{equation}
and 
\begin{equation}
\begin{array}{c}
\omega _{2}=\Omega D_{t}c_{eq}^{0}q^{2}\left\{ -\Gamma \left[ 
{\displaystyle {2\left( 1-\cos \phi \right)  \over l+2d}}%
+\left( l+%
{\displaystyle {s \over R}}%
\right) q^{2}\right] \right. \\ 
\left. +f\left( 
{\displaystyle {2dl \over l+2d}}%
+%
{\displaystyle {s \over R}}%
\right) \right\} ,
\end{array}
\label{eq:2Dwd_bcf_sum}
\end{equation}

$\omega _{1}$ characterizes the 1D instability and thus is independent of $q$.
The bunching instability occurs for $df>0$ with most unstable mode giving
step pairing with $\phi =\pi $. Note that Eq.\ (\ref{eq:bch_bcf_sum}) is
identical to Eq.\ (\ref{eq:bnch2r}), when Eq.\ (\ref{eq:mapk}) is used.

$\omega _{2}$ characterizes 2D wandering instabilities with respect to
perturbations of wavenumber $q$. The first term on the
right hand side is stabilizing, and has
its minimum value for $\phi =0$, where it is proportional to $\Gamma q^{4}$
and all the steps wander in phase.

The second term, proportional to the field,
contains two destabilizing contributions. The first contribution,
proportional to $D_{t}dfq^{2}$, describes a Mullins-Sekerka or Bales-Zangwill
instability induced by the terrace concentration gradient that can occur when
$df>0$.

The second contribution, proportional to $D_{s}sfq^{2}$, represents an alternative
mechanism for step wandering induced by field-driven
periphery diffusion along the step.
When $d>0$, both mechanisms operate with a step-down current, while the step-up
case is completely stable. When $d<0$, the second mechanism can produce
wandering with a step-down current, while bunching occurs for a step up
current, as was discussed earlier in Sec.\ (\ref{sec:stepwandering}).
These stability results are summarized in Table \ref{Table:LSR}.

\section{Implications for Si Surfaces}

Thus far, both step bunching and wandering instabilities have been analyzed
in general terms based on the simple idea of two-region diffusion. Now we
examine the implications for vicinal Si(111) surfaces.
If we assume for concreteness that reconstruction is
generally associated with slower adatom diffusion, we can give a
qualitatively reasonable scenario that can account for many features of the
electromigration experiments observed on Si(111).

In temperature range I, we assume there exists reconstruction in both step
and terrace regions. Consistent with the analysis of Liu and Weeks, we
assume that at low temperature the adatom diffusion in the reconstructed
step region is slower than in the terrace region, i.e. $R>1$, corresponding
to cases (i) and (ii) in Fig.\ \ref{FIG:chemgrad_exp}. A step-down current
induces both step bunching and step wandering of Mullins-Sekerka type.
However the wandering is likely suppressed by the bunching instability. A
step-up current produces a stable uniform step train.

At higher temperatures, we expect reconstruction in step region could have a
more fragile structure when compared to that in the terrace region since
step atoms have more dangling bonds. Thus there could exist an intermediate
temperature range where because of changes in the step reconstruction,
diffusion is faster in the step region than on terraces, i.e. $R<1$,
corresponding to cases (iii) and (iv) in Fig.\ \ref{FIG:chemgrad_exp}. The
uniform step train now exhibits bunching with a step-up current. Wandering
occurs with a step-down current, induced by driven diffusion parallel to the
step. In particular, if we substitute in Eq.\ (\ref{def:xi}) the latest
experimental values for the step stiffness, \cite{Si111stifT2exp_Cohen}
$\tilde{\beta}=16.3meV/\AA $, and for the effective charge, \cite{note_vcheck}
$z^{*}=0.13$, and use a typical electric field strength of $E=7V/cm$, the 
resulting wavelength is roughly given by $\lambda \simeq 2\pi \xi \sim 5\mu m$,
comparable with experimental values
\cite{Si111wanexp_Degawa1,Si111wanexp_Degawa2,Si111wanexp_Degawa3} of $6-9\mu m$.

In this picture, the transition between different temperature regimes is
associated with local equilibrium where $R=1.$ Conceivably, such a
transition could happen again at higher temperatures, since only small
changes in the relative diffusion rates can take the fundamental parameter
$R$ from less than to greater than unity and vice versa. This scenario
provides a consistent interpretation of experiments in the second
temperature regime and suggests more generally why there could be such
a complicated temperature dependence.

\section{Nonlinear Evolution of Current-induced Instabilities}

\subsection{Velocity Function Formalism}

To calculate the long time morphology of vicinal surfaces, effective
equations relating the velocity of a step to the local terrace widths have
proved to be very useful. \cite{vfn_form} A simple example of such a
velocity function is given by Eq.\ (\ref{eq:vnbch}). The {\em extended
velocity function formalism} \cite{vfn_Liu,vfn_Weeks} takes into account
also the capillarity of steps (line tension effects) as well as step
repulsions, which are needed to prevent step overhangs as the initial
instabilities grow. Here we also incorporate a periphery diffusion term, the
sharp step analogue of the parallel diffusion flux in the two-region model.
Thus the general form of the velocity function can be written as: 
\begin{equation}
v_{n}(y)=f_{+}\left( l_{n};\mu _{n},\mu _{n+1}\right) +f_{-}\left(
l_{n-1};\mu _{n-1},\mu _{n}\right) -\partial _{\tau }{\bf J}_{s}
\label{eq:v_ext}
\end{equation}
where $l_{n}(y)$ is the local width of terrace $n$ that is in front of step $
n$ and $\mu _{n}(y)$ is the local chemical potential of the step $n$.

The velocity functions $f_{\pm }$ contains contributions both from driven
fluxes on the two neighboring terraces given by the sharp step equivalence
of Eq.\ (\ref{eq:vnbch}), and equilibrium relaxation terms that can be
calculated in terms of the step edge chemical potentials $\mu _{n}$. \cite
{Relaxbchthy_Liu} The $\mu _{n}$ take account of both capillary effects for
an individual step (using a linear approximation for the curvature) and the
effects of nearest neighbor step interactions as described earlier. See
Refs. \onlinecite{vfn_Liu} and \onlinecite{vfn_Weeks} for
detailed expressions for $f_{\pm }$ and $\mu _{n}$.


\begin{figure}[tbp]
\includegraphics[width=76mm,height=57mm]{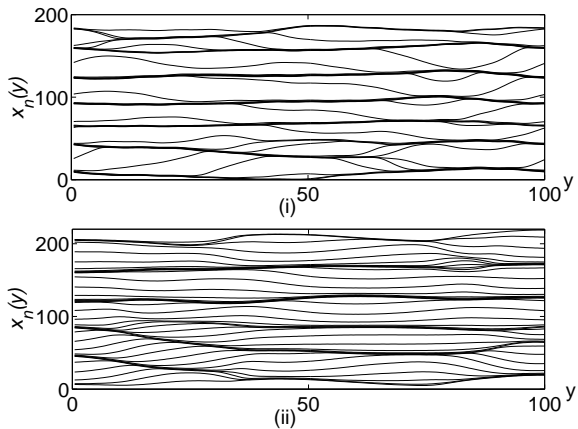}
\caption{A uniform step train composed of $30$ steps with spacing of 
$l=10$ forms step bunches at later times both for (i) $f>0,R>1$
and (ii) $f<0,R<1$.}
\label{FIG:bchcom}
\end{figure}

Numerically integrating Eq.\ (\ref{eq:v_ext}), we find step bunching patterns
for two parameter regimes (i) $f>0$, $R>1$ and (ii) $f<0$, $R<1$, in
agreement with predictions of linear stability analysis. The bunching
patterns in these two regimes are qualitatively similar, as shown in Fig.\ 
\ref{FIG:bchcom}. In both cases, step bunches form and
grow. In between the step bunches there are crossing steps traveling from
one bunch to the other.

\begin{figure}[tbp]
\includegraphics[width=76mm,height=57mm]{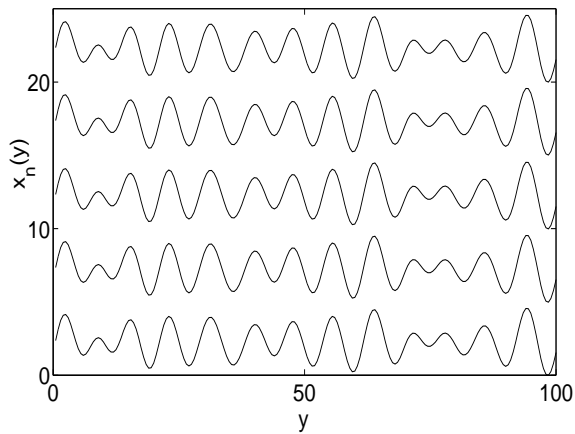}
\caption{A uniform step train comprised of $5$ steps with spacing
$l=5$ forms in-phase wandering patterns at later times for $f>0,R<1$.
Notice there are some defects in the pattern because the wandering
wavelength is incommensurate with the finite size
of our system in the $y$-direction.}
\label{FIG:vwanderm}
\end{figure}

In-phase step wandering is also given by Eq.\ (\ref{eq:v_ext})
in the regime $f>0$, $R<1$, as suggested by the previous linear
stability analysis. Typical wandering patterns with model parameters
are shown in Fig.\ \ref{FIG:vwanderm}.
Even though this is known to be a linear
instability, numerically we observe that it acts very like a nucleation
process. The steps fluctuate randomly as if the surface were completely
stable until a sinusoidal perturbation of the right wavelength forms. Once
formed, these small scale sinusoidal waves propagate through effective
``pulling'' by capillary effects in the lateral direction and by step
repulsions in the normal direction, until the entire surface is covered.
This is qualitatively consistent with experimental findings on Si(111).\cite
{Si111wanexp_Degawa3}

\subsection{Evolution of Step Wandering in a Geometric Representation}

Although Eq.\ (\ref{eq:v_ext}) has captured many physical features, it uses a
linearized curvature approximation and cannot be trusted when the step
curvature becomes large. Recent experiments show a continuous distortion of
the sinusoidal wandering wave by a field directed at an angle to the step
normal. We treat this problem here using a geometrical representation \cite
{GMthy1_Brower,GMthy2_Brower} of the step, where a single curve is 
parameterized by
intrinsic properties like its arc length $\tau $ and curvature $\kappa $.

It suffices to concentrate on a single step, since step wandering occurs in
phase. Consider a geometric representation of our step region with constant
width $s$, as in Fig.\ \ref{FIG:wanderschm}. The morphology of the step
region is specified by the position vector ${\bf x}\left( t,\tau \right) $
of the atomic step in the middle, where $\tau $ can represents the arc
length measured from an arbitrary origin. To follow ${\bf x}\left( t,\tau
\right) $ at a later time we need to know the velocity of the curve 
\begin{equation}
\frac{\partial {\bf x}}{\partial t}=v_{n}\hat{n}+v_{\tau }\hat{\tau},
\label{eq:v_gm0}
\end{equation}
where $\hat{n}$ and $\hat{\tau}$ denote normal and tangential directions as
before.

A general treatment of time-dependent curvilinear coordinates \cite
{BLMlong_Weeks} shows the equation of motion for the curve is 
\begin{equation}
\frac{\partial \kappa }{\partial t}=-\left[ \kappa ^{2}+\frac{\partial ^{2}}{
\partial ^{2}\tau }\right] v_{n}+v_{\tau }\frac{\partial \kappa }{\partial
\tau },  \label{eq:k_gm}
\end{equation}
which is subjected to the nonlocal metric constraint 
\begin{equation}
\frac{\partial \tau }{\partial t}=v_{\tau }\left( \tau \right) -v_{\tau
}\left( \tau =0\right) +\int^{\tau }v_{n}\kappa ds^{\prime }.
\label{eq:tau_gm}
\end{equation}
Interpreting $\tau $ as the arc length is arbitrary and other
parameterizations can be used, since only the normal velocity of the curve
is physically relevant. Following previous workers, \cite{GMthy2_Brower} we
take advantage of this ``gauge freedom'' and choose the {\em orthogonal gauge%
}, where $\tau $ is chosen at each instant of time so that the interface
velocity has only a normal component ($v_{\tau }=0$).

Now, we need to determine the normal velocity along the step. For simplicity, 
we will neglect contributions from the
terrace diffusion field as well as from the normal diffusion field in the
step region, since it has already been shown that the wandering instability
we are interested in is induced by the biased diffusion parallel to the step.
In the quasi-stationary limit, the diffusion field inside the step region is 
stationary for any given step position. To a good approximation, it can be 
taken as $c_{s}\simeq c_{s}^{0}\left( 1+\Gamma \kappa \right) $, where 
$c_{s}^{0}=c_{eq}^{0}s$ is the adatom density per unit step length for
straight steps. 

Next we consider the time rate of change of the adatoms 
contained in an element of the step region with an infinitesimal length 
$\delta \tau $ that moves with velocity $v_n$ as in Fig. \ref{FIG:wanderschm}.
This balance contains contributions from the motion of the step, and 
from the divergence of the flux parallel to the step. The latter accounts for 
diffusion driven both by the field and by chemical potential variations arising
from changes in step curvature. We thus have   
\begin{eqnarray}
\left[ \frac{d}{dt}\left( c_{s}\delta \tau \right) \right] _{n} &=&-\Omega
^{-1}v_{n}\delta \tau -D_{s}\partial _{\tau }\left[ fc_{s}\sin \left(
\varphi -\theta \right) \right] \delta \tau  \nonumber \\
&&+D_{s}\partial _{\tau }^{2}c_{s}\delta \tau .  \label{eq:drv_gm}
\end{eqnarray}
Using the exact geometrical relation $\left[ d\left( \delta
\tau \right) /dt\right] _{n}=v_{n}\kappa \delta \tau ,$ which can be
understood physically as the rate at which the arc length $\delta \tau $ on
a circle of radius $\left| \kappa ^{-1}\right| $ changes if the circle grows
only radially at rate $v_{n}$, Eq.\ (\ref{eq:drv_gm}) reduces to the
following form 
\begin{eqnarray}
\frac{v_{n}\left[ 1+\Omega c_{s}^{0}\left( 1+\Gamma \kappa \right) \kappa
\right] }{\Omega D_{s}c_{s}^{0}} &=&f\cos \left( \varphi -\theta \right)
\left( 1+\Gamma \kappa \right) \kappa  \nonumber \\
&&-f\sin \left( \varphi -\theta \right) \Gamma \partial _{\tau }\kappa
+\Gamma \partial _{\tau }^{2}\kappa .  \nonumber \\
&&  \label{eq:v_gm}
\end{eqnarray}
Combining Eq.\ (\ref{eq:v_gm}) with Eqs.\ (\ref{eq:k_gm}) and (\ref{eq:tau_gm})
yields a complete description of the dynamics of a single step region in
the presence of an electric field at an angle $\varphi $ off the $x$-axis.

\begin{figure}[tbp]
\includegraphics[width=76mm,height=57mm]{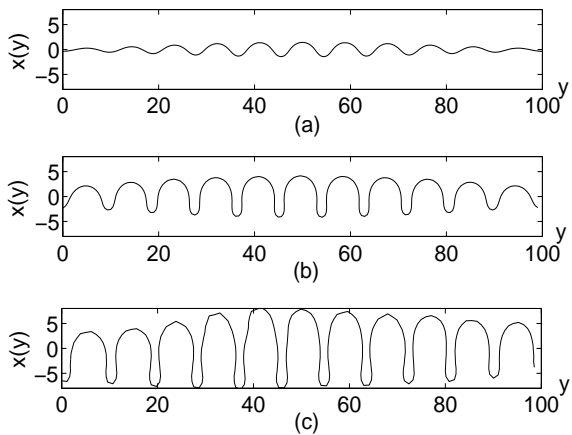}
\caption{Step evolution under a perpendicular electric field (a) At $t=160$,
a linear instability develops; (b) At $t=170$, asymmetry between the peaks and
valleys creates a periodic cellular structure; (c) At $t=190$,
the cellular shape is preserved but it grows in amplitude.}
\label{FIG:emgm90}
\end{figure}

\begin{figure}[tbp]
\includegraphics[width=76mm,height=57mm]{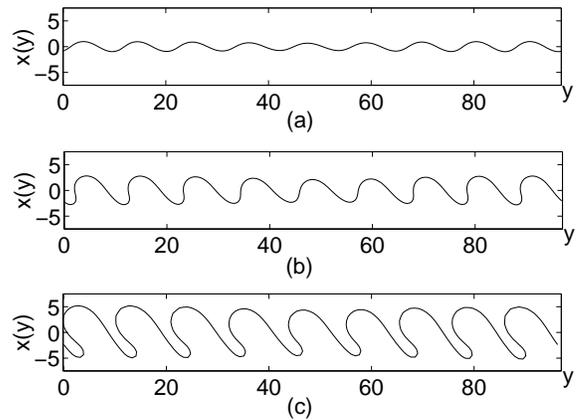}
\caption{Step evolution when the electric field is at an angle
$\varphi=\pi/4$ from the $x$-axis: (a) $t=300$, the initial instability induced by
the normal component of the field; (b) $t=315$, the peaks have begun to
turn; (c) $t=330$, all the peaks align with the direction of the field.}
\label{FIG:emgm45}
\end{figure}

We first consider the special case $\varphi =0$
where the external field is perpendicular to the average step direction
(the $y$ -axis). In Fig.\ \ref{FIG:emgm90}, we show three step configurations
evolving from a straight step with a small perturbation in the middle. The
linear wandering instability develops first as shown in Fig.\ \ref{FIG:emgm90}%
(a), then gradually changes into a cellular shape with the wavelength selected
by the linear instability, as illustrated in Fig.\ \ref{FIG:emgm90}(b). At
later stages, the cellular shape grows without significant distortion or
overlap, as shown in Fig.\ \ref{FIG:emgm90}(c). Notice that indeed we observe
numerically a long time period before the linear instability is significant.

\begin{figure}[tbp]
\includegraphics[width=76mm,height=57mm]{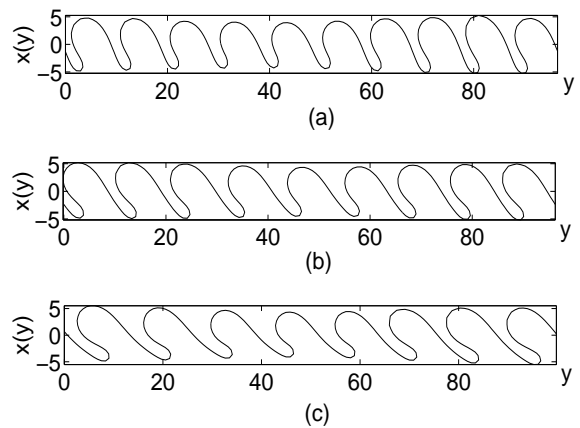}
\caption{Comparison of the step evolution as the angle $\varphi$ increases:
(a) $t=230$, $\varphi=\pi/6$; (b) $t=330$, $\varphi=\pi/4$; (c) $t=640$, $
\varphi=\pi/3$.}
\label{FIG:emgm_ang_comp}
\end{figure}

In Fig.\ \ref{FIG:emgm45}, we show configurations of the system with $\varphi =\pi
/4$. Fig. \ref{FIG:emgm45} suggests that the linear instability is
induced by the perpendicular component of the field. However, as the magnitude
of the instability grows, the peaks turn gradually until they are aligned
with the direction of the field. We see the same peak turning process when
the angle $\varphi $ is varied while keeping $f$ constant. However, since
the perpendicular component decreases with increasing $\varphi$, both the
wavelength selected by the initial instability as in Eq.\ (\ref{def:xi}) and
the time period before it forms increases monotonically with $\varphi $. The
numerical results for three particular angles are shown in Fig.\ \ref
{FIG:emgm_ang_comp}.

\begin{figure}[tbp]
\includegraphics[width=76mm,height=57mm]{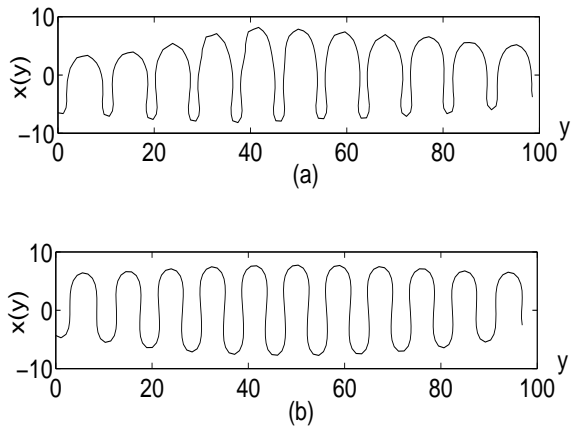}
\caption{A study of the asymmetry of the cellular patterns: (a) $t=180$, a snapshot of
the system given by Eq. (\ref{eq:theta0}). Note the close agreement with
Fig.\ \ref{FIG:emgm90}(c). This shows that the simplifed Eq.\ (\ref{eq:vapprox_gm})
with only terms linear in $\kappa $
captures most features of Eq. (\ref{eq:v_gm});
(b) $t=180$, a snapshot of a
model equation where the term 
$\sim\partial_{\tau}\kappa$ is left out of Eq.\ (\ref{eq:theta0}). Clearly this
term is mainly responsible for the asymmetric shape in (a).}
\label{FIG:emgm90_comp}
\end{figure}

To provide a more qualitative understanding of the pattern formation process,
we neglect the higher
order terms in $\kappa $ in Eq.\ (\ref{eq:v_gm}). To linear order in
$\kappa $, Eq.\ (\ref{eq:v_gm}) becomes 
\begin{equation}
\frac{v_{n}}{\Omega D_{s}c_{s}^{0}}=f\kappa \cos \left( \varphi -\theta
\right) -f\sin \left( \varphi -\theta \right) \partial _{\tau }\kappa
+\Gamma \partial _{\tau }^{2}\kappa .  \label{eq:vapprox_gm}
\end{equation}
In particular, for $\varphi =0$%
\begin{equation}
\frac{v_{n}}{\Omega D_{s}c_{s}^{0}}=f\kappa \cos \theta +f\sin \theta
\partial _{\tau }\kappa +\partial _{\tau }^{2}\kappa .  \label{eq:theta0}
\end{equation}

In the usual Mullins-Sekerka instability $\kappa $ alone appears in the
first term. Here however we have $\kappa \cos \theta$, resulting from
field driven diffusion inside the step region.
The extra $\cos \theta $ term brings in a field induced anisotropy that
makes the peaks and valleys of a perturbation preferably grow rather than
the sides. This stabilizes cellular structures. This anisotropy will keep
the tip unsplit, and it provides a cut off as the sides become nearly
vertical. Thus the cellular shapes formed under the influence of the
external field do not emit side branches, in contrast to most systems that
undergo a Mullins-Sekerka instability.

The second term in Eq.\ (\ref{eq:vapprox_gm}) is a flux induced by $-\kappa $
that effectively transports mass from the bottom to the top of a bulge and
is responsible for the asymmetric shape of the peaks and valleys, as is
illustrated in Fig.\ \ref{FIG:emgm90_comp}.

Although Eq.\ (\ref{eq:vapprox_gm}) is linear in the curvature,
$\kappa $ itself is a highly nonlinear function of the deviation from a
straight step. The early evolution is governed by the following linearized
equation 
\begin{equation}
\frac{1}{\Omega D_{s}c_{s}^{0}}\frac{\partial {x}}{\partial {t}}=-f\cos
\varphi \frac{\partial ^{2}{x}}{\partial y^{2}}-f\sin \varphi \frac{\partial
^{3}{x}}{\partial y^{3}}-\Gamma \frac{\partial ^{4}{x}}{\partial y^{4}}.
\label{eq:linear_in_x}
\end{equation}
The above equation is unstable when $f\cos \varphi >0$,
suggesting that the wavelength selection is determined by
the perpendicular component of the field. For $\varphi =0$, perturbations
with wavenumber $q_{0}=1/(\sqrt{2}\xi )$ are maximally amplified. For
$0<\varphi <\pi /2$, the most unstable wavenumber selected by the linear
instability is decreased by a factor of $\sqrt{\cos \varphi }$, i.e.,
$q_{\varphi }=q_{0}\sqrt{\cos \varphi }$.

As the instability grows, the field induced anisotropy characterized by the
factor $\cos (\varphi -\theta )$ becomes more significant. As in the $
\varphi =0$ case above, the anisotropy makes the initial sinusoidal wave
grows preferably in the direction where $\cos (\varphi -\theta )$ in Eq.\ (%
\ref{eq:vapprox_gm}) attains its minimum. Thus the wave will be continuously
distorted until the peaks point toward the field direction, and
subsequently only the magnitude of the pattern grows.

\section{Conclusion}

In this paper we have studied a physically suggestive two-region
diffusion model. The basic idea is to consider different hopping rates
associated with different reconstruction and bonding in the terrace and step
regions. The resulting steady state profiles provide important insight into
the physical origins of both step bunching and wandering instabilities. Step
bunching is induced by positive chemical potential gradients on terraces
that are essentially determined by the sign of $f(R-1)$. We argue that
step wandering in
Si(111) does not arise from the well known Mullins-Sekerka instability.
Rather, it is induced by
driven diffusion along the step edge under the influence of a step-down
force, and only becomes significant when step bunching is absent, which
requires a negative kinetic coefficient.

We also carried out a mapping from the two-region model to a sharp step
model using a simple extrapolation procedure. The result connects the
kinetic coefficients in sharp step models to relative diffusion rates in
terrace and step regions. In particular, the lowest order result shows that
the kinetic coefficients are independent of the driving field, in contrast
to earlier suggestions.\cite{Si111bchthy_Suga}

A coherent scenario for Si(111) electromigration is proposed based on the
stability analysis of the model. In particular, the mysterious second
temperature regime is interpreted using a negative kinetic coefficient. This
allows the step wandering that generally occurs with a step-down force to be
separated from step bunching. The transition between different temperature
regimes is governed by the relative diffusitivity in the terrace and step
regions. Other theories can predict a reversal of step bunching arising from
a change in step transparency \cite{Si111bchthy_Stoyanov,Si111bchthy_Sato}
or from a change of sign of the effective charge. \cite{Effcharge_Kandel}
However, neither approach can give a consistent treatment for step wandering.

The long time evolution of the step instabilities was calculated by
numerical integration of a set of equations based on the standard velocity
function formalism with the addition of a periphery diffusion term. The
linear instabilities are recovered at short times and interesting 2D pattern
formation is see at longer times in qualitative agreement with experiment.

We also showed that a geometric representation of the step provides a
simple way to describe the nonlinear evolution of step wandering
patterns with large curvatures. The resulting cellular patterns when the
driving field is at an angle to the step shows significant step
``overhangs'', which can not be captured by standard multi-scale expansion
methods. \cite{Multiscale_Bena,Multiscale_PierreL}

The two-region model can also be modified to explain many features of the
very different step bunching behavior seen on Si(001) surfaces. \cite
{hopping_Zhao} Thus it provides a simple and unified perspective that can
shed light on both general properties of current-induced step bunching and
wandering instabilities and their specific manifestations on Si surfaces.

\begin{acknowledgments}
We are especially grateful to Daniel Kandel for many stimulating discussions 
on connecting the two-region model to sharp step models.
We also thank Ted Einstein, Oliver Pierre-Louis, and 
Ellen Williams for helpful comments. This work has been supported by the 
NSF-MRSEC at the University of Maryland under Grant No. DMR 00-80008.
\end{acknowledgments}

\appendix*

\section{Linear Stability Analysis in a Sharp Interface Model}

A complete 2D stability analysis in a generalized BCF model is performed in
this section, with boundary conditions dictated by mapping from the
two-region model. Using the quasi-stationary approximation, we first solve
for the static concentration field $c_{t}$ on the terrace as given by 
\begin{equation}
D_{t}\nabla ^{2}c_{t}-\frac{D_{t}{\bf F}}{k_{B}T}\cdot \nabla c_{t}=0,
\label{eq:diff_bcf}
\end{equation}
subject to the general linear kinetics boundary condition at the sharp step: 
\begin{equation}
\pm D_{t}\left[ \nabla c_{t}-fc_{t}\right] _{\pm }\cdot \hat{n}=k\left[
c_{t}-c_{eq}^{0}\left( 1+\Gamma \kappa \right) \right] _{\pm }.
\label{eq:bc_bcf}
\end{equation}
Here $\kappa $ is the curvature, defined to be positive for a circle. The
normal step velocity is determined by balancing the fluxes locally at the
step 
\begin{equation}
v_{n}\Delta c=\hat{n}\cdot \left[ {\bf J}_{t}^{-}-{\bf J}_{t}^{+}\right]
-\partial _{\tau }J_{s}.  \label{eq:vn_bcf}
\end{equation}
Here $J_{s}$ is the periphery flux of the mobile atoms along the interface,
which can be viewed as the coarse-grained contribution from the parallel
diffusion in the two-region model. In general, $J_{s}$ takes the form 
\begin{equation}
J_{s}=-D_{s}\partial _{\tau }c_{s}+D_{s}\frac{{\bf F}\cdot \hat{\tau}}{k_{B}T%
}c_{s},  \label{eq:jst_bcf}
\end{equation}
where $c_{s}\simeq c_{eq}^{0}s$ gives the effective number of ledge atoms per
unit step length.

Consider a 2D perturbation on the step profile in the form $\delta
x_{n}\left( y,t\right) \equiv x_{n}\left( y,t\right) -x_{n}^{0}=\varepsilon
_{n}e^{\omega t+iqy}+c.c.$, where $x_{n}^{0}$ is the step position for 1D
steady state and $\varepsilon _{n}$ is the 1D perturbation previously
defined. In general $\omega $ can be complex, i.e. $\omega =\omega
_{r}+i\omega _{i}$, but we are only interested in the real part $\omega _{r}$
whose sign determines the instability. The calculation follows standard
methods, and the result can cast in the familiar Bales and Zangwill's form 
\cite{MBEthy_Bales}: 
\begin{equation}
\omega _{r}=-\Gamma q^{2}h\left( q,f,\phi \right) +fg\left( q,f,\phi \right)
.  \label{eq:omegagen_bcf}
\end{equation}
Here the stabilizing piece $h\left( q,f,\phi \right) $ is given by 
\begin{eqnarray}
\frac{h\left( q,f,\phi \right) }{\Omega D_{t}c_{eq}^{0}} &=&\frac{2\lambda
\left[ \cosh \left( \lambda l\right) -\cos \phi \cosh \left( fl/2\right)
\right] +2dq^{2}\sinh (\lambda l)}{{\cal D}}  \nonumber \\
&&+\frac{a}{R}q^{2},  \label{eq:h_bcf}
\end{eqnarray}
and the destabilizing piece $g\left( q,f,\phi \right) $ is 
\begin{equation}
\begin{array}[b]{ll}
{\displaystyle {g\left( q,f,\phi \right)  \over \Omega D_{t}c_{eq}^{0}}}%
= & 
{\displaystyle {2df \over \left[ df\left( e^{fl}+1\right) +e^{fl}-1\right] {\cal D}}}%
\\ 
& \times \left\{ 2\lambda \left[ \cosh \left( \lambda l\right) -e^{fl/2}\cos
\phi \right] \right.  \\ 
& +2dq^{2}e^{fl/2}\cosh \left( fl/2\right) \sinh \left( \lambda l\right)  \\ 
& +\left. \sinh \left( fl/2\right) \left( \Lambda _{+}e^{\Lambda
_{-}l}-\Lambda _{-}e^{\Lambda _{+}l}\right) \right\}  \\ 
& +%
{\displaystyle {a \over R}}%
q^{2},
\end{array}
\label{eq:g_bcf}
\end{equation}
where 
\begin{equation}
{\cal D}=2d\lambda \cosh \left( \lambda l\right) +\left( 1+d^{2}q^{2}\right)
\sinh \left( \lambda l\right) ,  \label{eq:D}
\end{equation}
$\lambda =\sqrt{f^{2}+4q^{2}}/2$ and $\Lambda _{\pm }=f/2\pm \lambda $.

It is easy to see that $h\left( q,f,\phi \right) $ is positive definite;
thus the first term in Eq.(\ref{eq:omegagen_bcf}) is always stabilizing. In
particular, we obtain the results for the equilibrium relaxation by taking
the limit $f\rightarrow 0$ 
\begin{eqnarray}
\frac{\omega _{0}\left( q,\phi \right) }{\Omega D_{t}c_{eq}^{0}} &=&-\Gamma
q^{2}\left\{ \frac{2q\left[ \cosh \left( ql\right) -\cos \phi +dq\sinh
\left( ql\right) \right] }{2dq\cosh \left( ql\right) +\left(
1+d^{2}q^{2}\right) \sinh \left( ql\right) }\right.   \nonumber \\
&&+\left. 
{\displaystyle {s \over R}}%
q^{2}\right\} .  \label{eq:eqrestor_bcf}
\end{eqnarray}
The two terms in the curly brackets account for relaxation through terrace
diffusion and periphery diffusion respectively. The second term in Eq.\ (\ref
{eq:omegagen_bcf}) is completely driven by the field, and it vanishes
identically as $f\rightarrow 0$ since $\lim_{f\rightarrow 0}g\left( q,f,\phi
\right) $ is finite.


For current-induced instabilities that are of interest of this paper, we can
take the weak field $\left( fl\ll 1\right) $ and long wavelength $\left(
ql\ll 1\right) $ limit. In this limit, the stability functions to linear
order in the field are given in Eqs.\ (\ref{eq:stblygen_bcf_sum})-(\ref
{eq:2Dwd_bcf_sum}).

\end{document}